\documentclass[lettersize,journal]{IEEEtran}
\usepackage{amsmath,amsfonts}
\usepackage{algorithmic}
\usepackage{algorithm}
\usepackage{array}
\usepackage[caption=false,font=normalsize,labelfont=sf,textfont=sf]{subfig}
\usepackage{textcomp}
\usepackage{stfloats}
\usepackage{url}
\usepackage{verbatim}
\usepackage{graphicx}
\usepackage{cite}
\begin{document}

\title{Power-dependent Reflective Metasurface with Self-induced Bandgap}

\author{Xiaozhen Yang, Erda Wen and Daniel F. Sievenpiper,~\IEEEmembership{Fellow,~IEEE}
\thanks{This work is supported by Office of Naval Research under Grant No. N0014-20-1-2710.}
\thanks{The authors are with the Electrical and Computer Engineering Department,
University of California San Diego, La Jolla, CA 92093-0021 USA (e-mail:
xiy003@eng.ucsd.edu).}
}


\maketitle

\begin{abstract}
A metallic ring based, diode-integrated, low-profile, power-dependent, reflective metasurface working from $3$ GHz to $3.6$ GHz is proposed in this letter. Unlike the previous study which shifts a band up and down to change the impedance of the surface, the triggering of the diodes directly transforms the structure from a surface wave supportive state to a self-induced bandgap topology if exposed to high power RF illumination. We demonstrate the concept by conducting the EM-circuit co-simulation and measurements for a $6$ by $8$ unit 2D prototype. Near field scan experiments verify that the proposed topology works in two distinct states, the ON and OFF state, and high-power measurements prove that the reflection varies with the incident signal power. The highest $10$ dB decrement in transmission occurs at $3.3$ GHz with $52$ dBm illumination. This structure can be used to protect sensitive devices from large signals while otherwise supporting a communication channel for small signals.
\end{abstract}

\begin{IEEEkeywords}
Metasurface, interference suppression, high power, diodes
\end{IEEEkeywords}

\section{Introduction}
\IEEEPARstart{T}{he} development of microwave technology \cite{Collin07_FUND,Pozar11_FUND} in electronic devices and communication systems has brought much convenience to daily life. However, such low-power and sensitive systems can be easily interfered, or even damaged, by the complex electromagnetic environment or by external microwave sources. The incident waves can induce currents, which also radiate into the shielding, around any gaps in a metallic enclosure, even if the openings are electrically small. 

Microwave absorbers are used to protect devices from incoming signals by absorbing the energy. Conventional approaches \cite{Ruck_70,Knott_12} such as Salisbury screen \cite{Salisbury_52} apply lossy coatings, which are usually bulky and expensive, to the surface of the metallic enclosure to dissipate the power. Recent research on metasurfaces offer another solution\cite{Sievenpiper_99,Wang_20,Landy_08}. These low-profile, low-cost, sub-wavelength, PCB-based, periodic metallic structures can be engineered to various frequency ranges with adequate performance. 

Although these approaches could reduce the transmission through gaps and openings, such linear absorption behaviour of conventional absorbers can also impair the performance of the protected device itself, especially for low-power communication systems. Nonlinear \cite{Wakatsuchi_13}, switchable \cite{Sievenpiper_11,Li_17,Kim_16} and tunable \cite{Xiong_20,Song_19,Mou_20} meatsurface absorbers have arisen attention in recent years. By integrating with lumped elements, diodes, switches, and transistors, the absorbers work in different states dependent on the power level \cite{Luo_15,Kim_16} or the waveform \cite{Wakatsuchi_13_2} of the incoming signals due to the nonlinearity provided by the elements. However, these structures are usually complex and some require control circuits.

\begin{figure}[!t]
\centering
\includegraphics[width=3.5in]{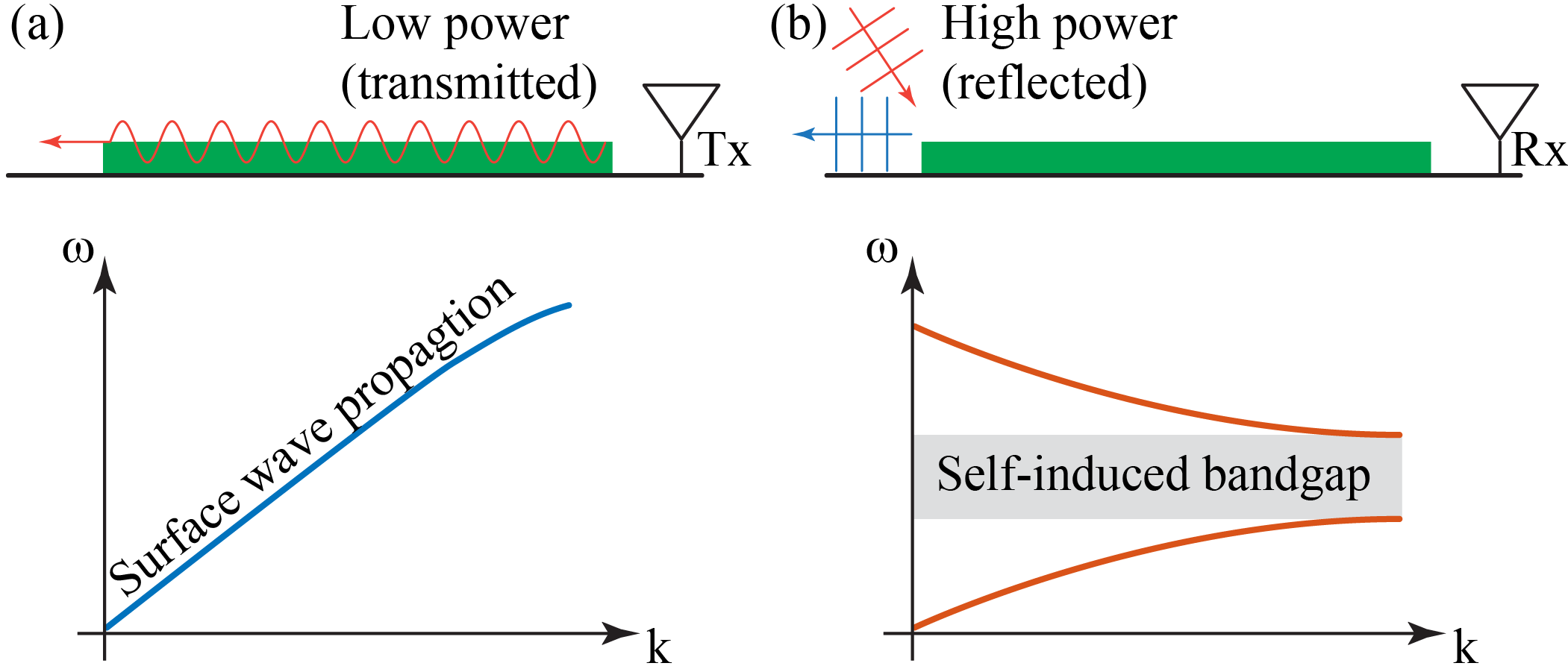}
\caption{Nonlinear reflective surface behaviour and its band structure. (a) Low power illumination. (b) High power illumination.}
\label{concept} 
\end{figure}

In this letter, we propose a self-induced, diode-integrated, power-dependent, reflective metasurface which changes the topology of its surface wave band structure to create a self-induced bandgap with high power illumination as shown in Fig. \ref{concept}. Thus, this nonlinear surface reflects the incoming wave rather than absorbing it within the bandgap region. Unlike the previous study \cite{Kim_16}, rather than shifting the band up and down to change the surface impedance, and thus change the absorption rate, we directly manipulate the band structure to open a bandgap which prevents propagation of surface waves.

\begin{figure}[t]
\centering
\includegraphics[width=3.5in]{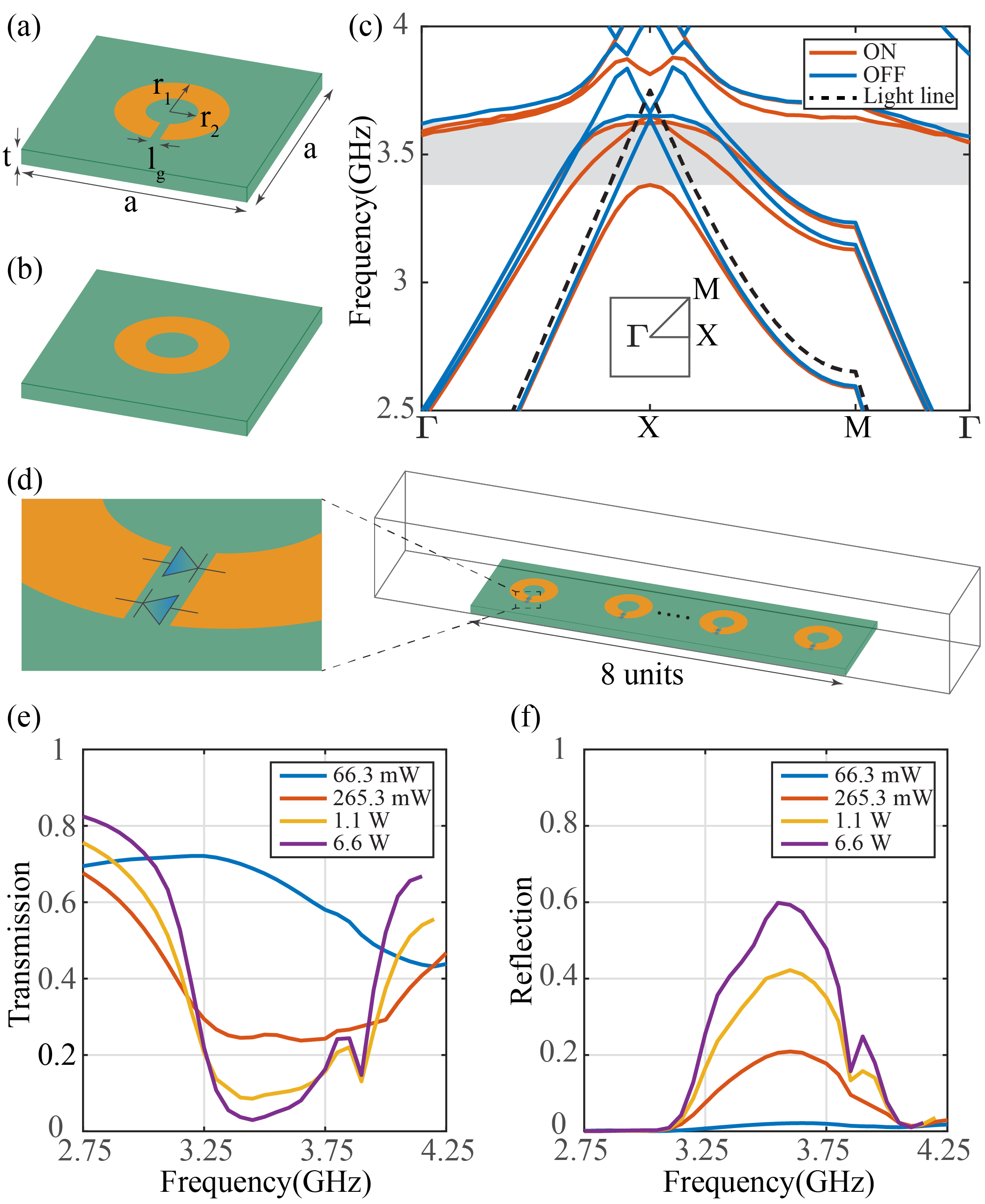}
\caption{The ON and OFF unit cell design and EM-co simulation results. (a) The OFF state unit cell design. $a=40$ mm, $r_{1}=9.6$ mm, $r_{2}=4.44$ mm, $l_{g}=1.7$ mm ,$t=3.175$ mm. The substrate is FR4 with $\epsilon_{r}=4.4$. (b) The ON state unit cell design. (c) Band diagram for the ON and OFF state. Shaded area: the first bandgap region. (d) Full wave simulation model in a TEM waveguide. (e) The transmission rate at different power levels simulated for the model in (d). (f) The reflection rate at different power level simulated for (d).}
\label{unit} 
\end{figure}

\section{Structure Design and Simulation Results}
\subsection{Static Unit Cell Design}
The unit cell is designed to transition between two states (the ON and OFF states) based on the incident power: the low power state (the OFF state) as shown in Fig.\ref{unit}(a) and the high power state (the ON state) as shown in Fig.\ref{unit}(b). Either a ``C'' or ``O'' shaped metallic ring is printed on a $3.175$ mm thick, $4$ cm by $4$ cm FR4 square lattice for the OFF and ON states, respectively, with a ground plane at the bottom. An eigenmode simulation is performed in an $8$ cm high airbox with PEC assigned on the top surface and periodic boundaries around it using Ansys Electronics Desktop. The band diagram is shown in Fig. \ref{unit} (c). The OFF state supports surface waves from $2.5$ GHz to $4$ GHz shown as the blue solid line. A complete bandgap below the light line, covering from $3.38$ GHz to $3.64$ GHz, is created for the ON state represented by the red solid line. Thus, this structure can be used within the bandgap region by integrating with diodes at the gaps of the ``C'' shape geometry as shown in Fig. \ref{unit} (d). When the incoming power is high, the gaps are closed by the triggered diodes, and thus the unit cells are in the ON state with a bandgap. If the illumination is not high enough to turn on the diodes, the metallic ring remains in the ``C'' shape, and the surface supports surface wave propagation throughout the entire frequency band.

\subsection{EM-Circuit Co-Simulation}
To prove this concept, an EM-circuit co-simulation is conducted using Ansys Electronic Desktop. The full wave simulation model is shown in Fig. \ref{unit} (d). The OFF state unit cell is expanded to an $8$ unit cell model placed inside a transverse electromagnetic (TEM) waveguide, whose top and bottom sides are assigned with PEC, front and back are assigned with PMC, and left and right are assigned with wave ports with proper characteristic impedance according to the size of the TEM waveguide. After solving for the S-matrix from DC to $12$ GHz, the lumped ports at the gaps are connected with a pair of oppositely oriented diodes to perform a transient simulation in the circuit simulator. One of the wave ports is connected with a sinusoidal voltage source at different power levels, and the other one is connected with a resistor with the same resistance as the characteristic impedance of the wave port to measure the transmission. The structure works in a hybrid state of the ON and OFF states depending on the incident power as shown in Fig. \ref{unit} (e) and (f). It is observed that as the power rises, the transmission decreases and the reflection increases for frequencies inside the bandgap range, proving that the structure can reduce transmission by reflection with a self-induced bandgap. At $3.45$ GHz, $70\%$ of the incident power propagates through the TEM waveguide with $66$ mW input, while only $3\%$ is collected by the receiving port with $6.6$ W input with $60\%$ of incident power reflected back to the transmitting port. Note that the rest of the power is absorbed by the lossy FR4 substrate (loss tangent $0.02$), which could be reduced by using a lower loss substrate such as Rogers Duroid 5880. The results show that the working band ($3$ GHz to around $4$ GHz) is wider than the bandgap shown in Fig. \ref{unit}(c), which stems from a second bangap above $3.64$ GHz.

\section{Measurements}
\begin{figure}[!t]
\centering
\includegraphics[width=3.5in]{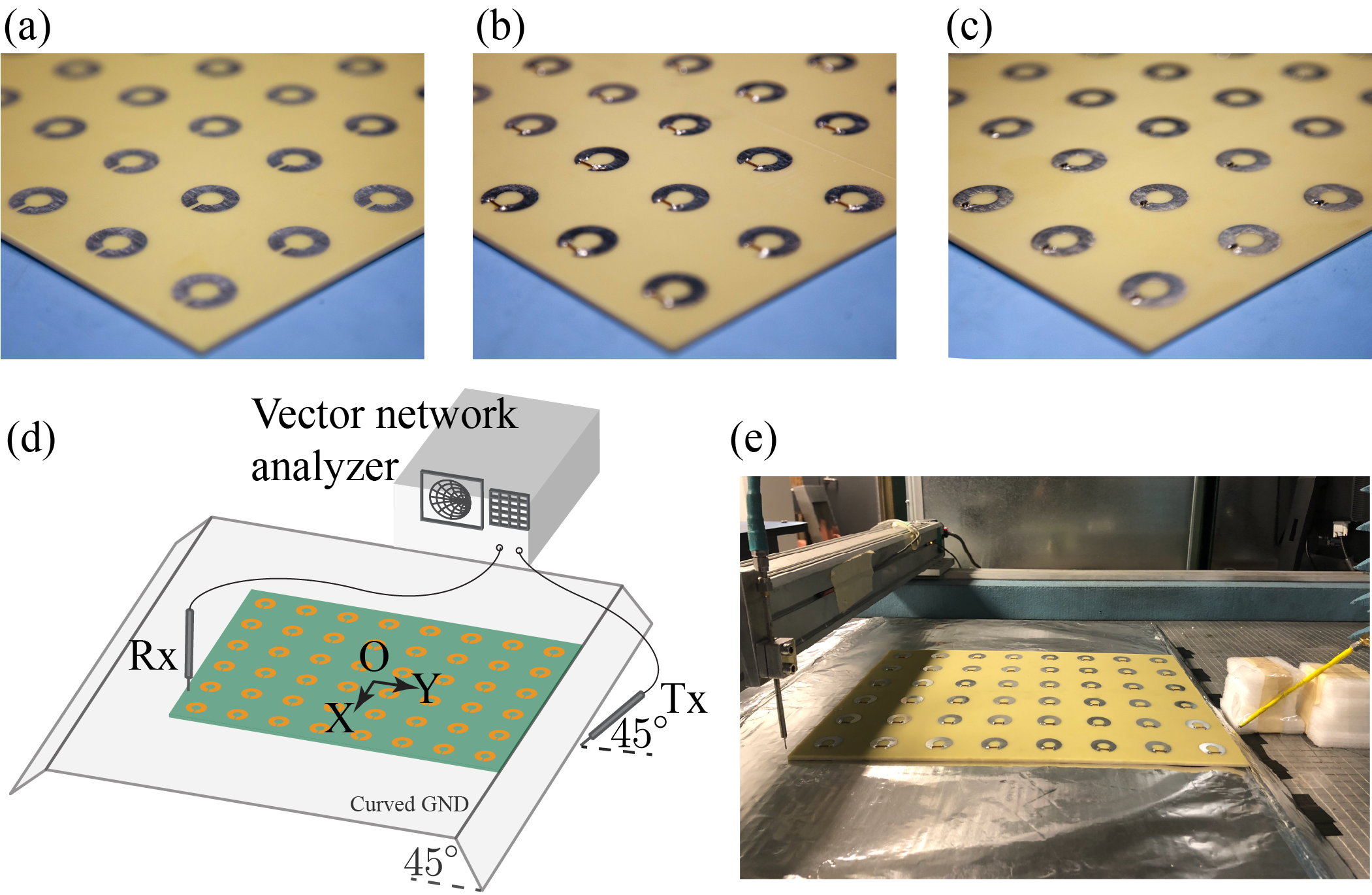}
\caption{(a) The OFF state prototype. (b) The ON state prototype. (c) The diode-integrated prototype. (d) The near field scan measurement setup. (e) A picture of near field scan measurement.}
\label{NFsetup} 
\end{figure}

To verify the existence of the bandgap and the power-dependent effect, two measurements are conducted on a pair of 2D $6$ by $8$ unit cell static surface and the diode PCB prototypes. All PCBs are manufactured in the OFF state. Copper wires are soldered at the gaps of the OFF state sample to create an ON state prototype in Fig. \ref{NFsetup}(b), and a pair of opposite oriented Schottky diodes (Infineon BAT15-04W) are connected at the gaps to build a diode-integrated reflective metasurface in Fig. \ref{NFsetup}(c). 

\subsection{Near Field Scan}
\begin{figure*}
\includegraphics[width=7.16in]{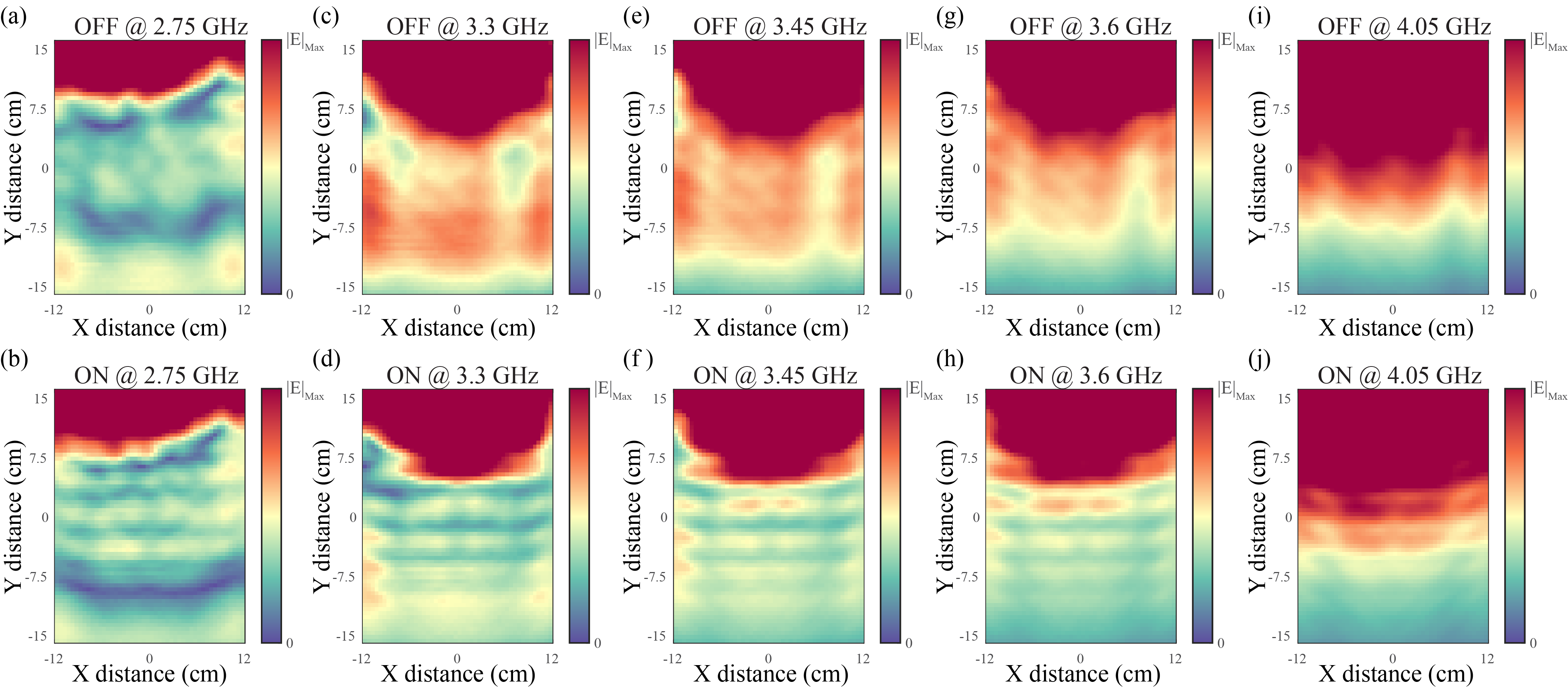}
\caption{The near field scan results for the ON and OFF states at different frequencies. The first row: the OFF state field scan after adopting time gating at $2.75$ GHz, $3.3$ GHz, $3.45$ GHz, $3.6$ GHz, $4.05$ GHz, respectively. The second row: the ON state field scan after adopting time gating at $2.75$ GHz, $3.3$ GHz, $3.45$ GHz, $3.6$ GHz, $4.05$ GHz, respectively.}
\label{NFresults} 
\end{figure*}

The first measurement is the near field scan, as shown in Fig. \ref{NFsetup}(d) and (e), to confirm the bandgap region of the manufactured samples. The prototype is placed on a lifted ground plane, the end of which is a $5$ cm long $45^{\circ}$ slope, to separate the surface wave coupled to the structure and the free space plane wave generated from the $45^{\circ}$ tilted transmitting probe Tx which is connected to a two-port vector network analyzer (Keysight E5071C). After travelling through the device under test, the signal is collected by the receiving probe Rx, which scans over the sample with a $0.5$ cm resolution in both X and Y directions at $1$ cm height above the ON and OFF boards. The transmission $S_{21}$ (average of 8 sets of measurements) from $2.5$ GHz to $4$ GHz is recorded to represent the $E_z$ field intensity. To filter out reflections from the surrounding environment, a $1$ ns time gating is applied in time domain by performing the inverse Fourier transform. The field scan at different frequencies are shown in Fig. \ref{NFresults}. 

Here we observe the bandgap from $3$ GHz to $3.6$ GHz. The frequency shift compared to the simulation results could be caused by the variation of the permittivity of the FR4 substrate. For frequencies outside the bandgap, such as $2.75$ GHz and $4.05$ GHz, the difference in the field scan results between the ON and OFF states is negligible. On the other hand, for frequencies inside the bandgap, such as $3.3$ GHz, $3.45$ GHz and $3.6$ GHz, the OFF state scan demonstrates a much stronger field compared to the ON state. These near field scan results indicate that the bandgap topology helps to reduce the transmission.

\subsection{High-power Measurement}

\begin{figure}[!t]
\centering
\includegraphics[width=3.5in]{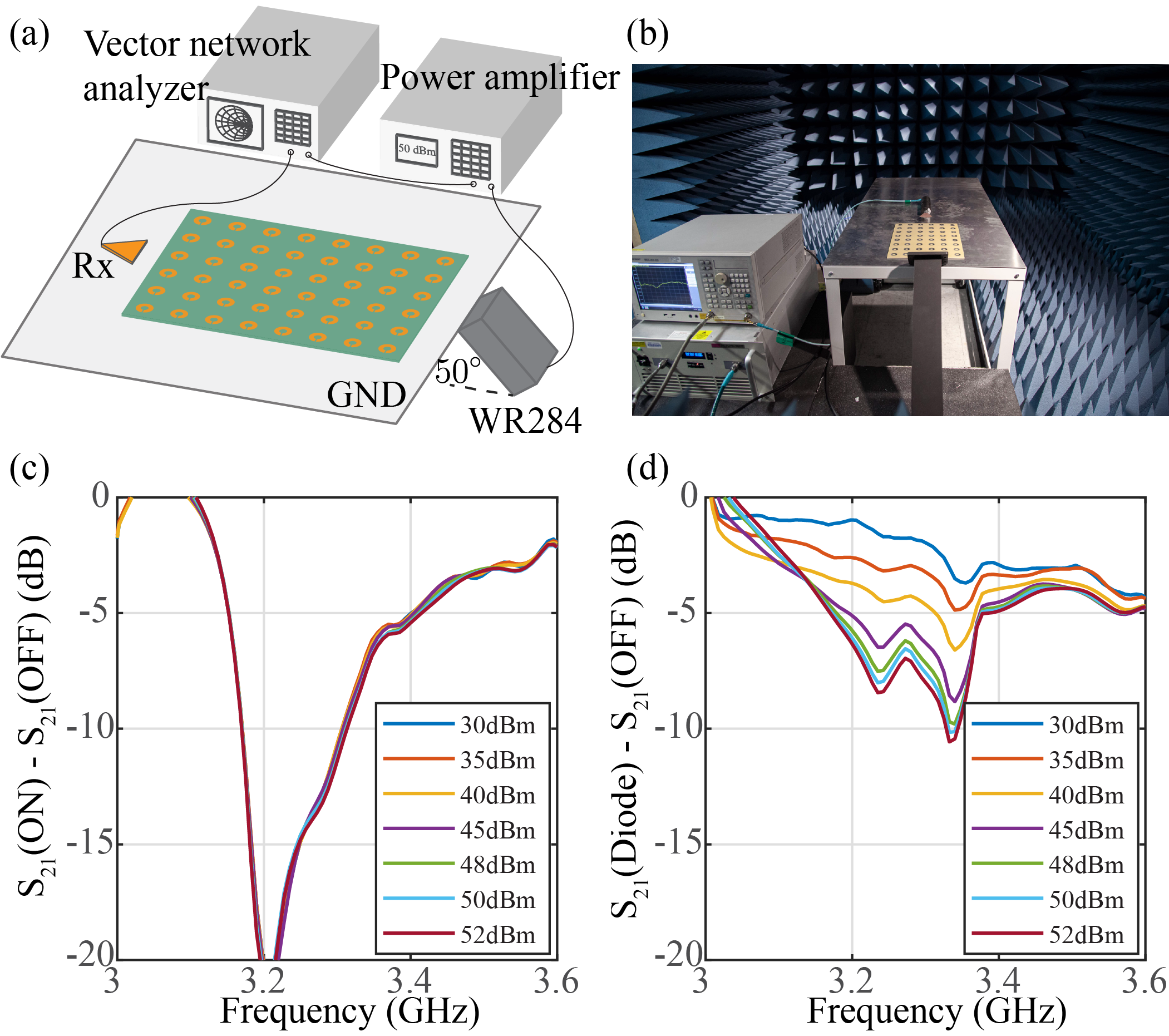}
\caption{The high-power measurement and results. (a) The measurement setup. (b) A picture of the measurement. (c) The static state results. (d) The diode-integrated prototype results.}
\label{HighPower} 
\end{figure}

After verifying the working status of the static state models, the diode-integrated reflective surface is measured under different power levels. The samples are placed on a $120$ cm by $60$ cm ground plane. As shown in Fig. \ref{HighPower}(a) and (b), the signal is generated by a VNA (Keysight E5071C) and amplified by an RF power amplifier (Ophir 5265), then radiated by a $50^{\circ}$ tilted rectangular waveguide (WR284, $2.6$ GHz to $3.95$ GHz) to separate the surface wave and the plane wave. The output power varied from $30$ dBm to $52$ dBm across $3$ GHz to $3.6$ GHz. After propagating through the device under test, the signal is received by a dielectric horn antenna Rx at $1$ cm height above the surface with a $40$ dB attenuator to guarantee the captured power is within the safe range of the VNA. This experiment is performed inside a sealed anechoic chamber, and all the measurement devices involved are remotely controlled to avoid human body exposure to RF radiation. To reduce the noise and fluctuations, the data is the average of $8$ measurements.

The static state prototypes, as well as the diode-integrated sample are measured under various power levels. The OFF state results are used as the baseline representing the surface with no effect on the incoming wave. Subtracting the measured $S_{21}$ for the OFF state from that for the ON state or diode-integrate sample, the difference directly indicates the impact brought by the self-induced bandgap. As shown in Fig. \ref{HighPower}(c), the difference between the ON and OFF state remains the same for different power levels across the whole frequency range. The largest difference occurs at $3.2$ GHz for all power levels. In contrast, the diode-integrated sample presents a power-dependent behaviour. As the incident power increases, the transmission decreases, reaching a $10.6$ dB drop at $3.33$ GHz with a $52$ dBm input. The difference in $S_{21}$ continuously increases as the power rises, because the diode gradually saturates and the reflection approaches a constant rate, showing a similar trend to the simulated results. The results shown in Fig.\ref{HighPower} (d) differs from that in (c) because the incident wave cannot trigger all the diodes, instead, only the diodes close to the Tx. Additionally, the parasitic capacitance of the diodes shifts the lowest transmission point from $3.2$ GHz to $3.33$ GHz.

\section{Conclusions}
 
In this letter, we propose the first power-dependent reflective metasurface by directly changing the band structure to open a bandgap under high power illumination. It automatically changes from a surface wave supportive state to a bandgap topology as the incident power increases. Measurements, together with simulation results, prove the validity of this concept. The near field scan shows the static behaviour of the proposed topology in the ON and OFF states, proving existence of a bandgap, which reflects incident signals, from $3$ GHz to $3.6$ GHz. The following high-power measurement demonstrates the power-level dependent performance for the diode-integrated prototype. The proposed sample shows $10$ dB less transmission compared to the OFF state at approximately $3.3$ GHz with $52$ dBm input.

The proposed structure is low-cost, low-profile and it has a simple geometry which does not require any control circuit. Additionally, it shows a wider transmission variation range ($10$ dB) compared to previous work on power-dependent absorbers (approximately $4.8$ dB for \cite{Li_17,Kim_16}). This reflective metasurface can be adopted to reduce RF exposure to sensitive devices while still allowing small signal communication for the protected systems. This self-induced bandgap concept opens a way to many similar designs based on manipulating the topology of the surface wave band structure.

\end{document}